%% file: Muzammel_MLWA.tex
\documentclass[a4paper,fleqn]{cas-dc}
\usepackage[authoryear]{natbib}
\usepackage{import}
\usepackage{microtype}
\usepackage{amssymb}
\usepackage{latexsym}
\usepackage{url}
\usepackage{subcaption}
\usepackage{pgfplots}
\usepackage[font=footnotesize,labelfont=bf]{caption}
\usepackage[font=footnotesize,labelfont=bf]{subcaption}
\usepackage{adjustbox}
\usepackage{graphicx}
\usepackage{xcolor}
\usepackage{array}
\usepackage{multirow}
\usepackage{tikz}
\usetikzlibrary{positioning}

\newcommand\MyBox[2]{
  \fbox{\lower0.75cm
    \vbox to 1.7cm{\vfil
      \hbox to 1.7cm{\hfil\parbox{1.4cm}{#1\\#2}\hfil}
      \vfil}%
  }%
}
\newcommand\MyBoxx[2]{
  \mbox{\lower0.75cm
    \vbox to 1.7cm{\vfil
      \hbox to 1.7cm{\hfil\parbox{1.4cm}{#1\\#2}\hfil}
      \vfil}%
  }%
}

\graphicspath{{figures/}}

\begin{document}
\let\WriteBookmarks\relax
\def\floatpagepagefraction{1}
\def\textpagefraction{.001}
\shorttitle{AudVowelConsNet: A Phoneme-Level Based Deep CNN Architecture for Clinical Depression Diagnosis}
\shortauthors{Muhammad Muzammel et~al.}

\title [mode = title]{AudVowelConsNet: A Phoneme-Level Based Deep CNN Architecture for Clinical Depression Diagnosis}

\author[1]{Muhammad Muzammel}%[type=editor,
                        %auid=000,bioid=1,
                        %prefix=Sir,
                        %role=Researcher,
                        %orcid=0000-0001-7511-2910]
%\cormark[1]
%\fnmark[1]
\ead{muhammad.muzammel@u-pec.fr}
%\ead[url]{www.cvr.cc, cvr@sayahna.org}

\author[2]{Hanan Salam}
%\cormark[2]
%\fnmark[2]
\ead{salam@em-lyon.com}
%\ead[URL]{www.hanansalam.com}

\author[3]{Yann Hoffmann}
%\cormark[3]
%\fnmark[3]
\ead{yann.hoffmann.002@student.uni.lu}
%\ead[URL]{www.stmdocs.in}

\author[4]{Mohamed Chetouani}
%\cormark[4]
%\fnmark[4]
\ead{mohamed.chetouani@sorbonne-universite.fr}
%\ead[URL]{www.stmdocs.in}

\author[1]{Alice Othmani}
\cormark[1]
%\fnmark[1]
%\ead{alice.othmani@u-pec.fr}
%\ead[URL]{www.stmdocs.in}

\address[1]{Université Paris-Est Créteil (UPEC), LISSI, Vitry sur Seine 94400, France}
\address[2]{Emlyon, 23 Avenue Guy de Collongue, 69130 \'Ecully, France}
\address[3]{University of Luxembourg, Kirchberg Campus, 6, rue Richard Coudenhove-Kalergi, L-1359, Luxembourg}
\address[4]{Institut des Systèmes Intelligents et de Robotique, Sorbonne Universite, CNRS UMR 7222, Paris, France}

\cortext[cor1]{Corresponding author: Dr. Alice OTHMANI \ead{alice.othmani@u-pec.fr}}

\begin{abstract}
Depression is a common and serious mood disorder that negatively affects the patient's capacity of functioning normally in daily tasks.
Speech is proven to be a vigorous tool in depression diagnosis. 
Research in psychiatry concentrated on performing fine-grained analysis on word-level speech components contributing  to the manifestation of depression in speech and revealed significant variations at the phoneme-level in depressed speech. On the other hand, research in Machine Learning-based automatic recognition of depression from speech focused on the exploration of various acoustic features for the  detection of depression and its severity level. Few have focused on incorporating phoneme-level speech components in automatic assessment systems. In this paper, we propose an Artificial Intelligence (AI) based application for clinical depression recognition and assessment from speech. We investigate the acoustic characteristics of phoneme units, specifically vowels and consonants for depression recognition via Deep Learning. We present and compare three spectrogram-based Deep Neural Network architectures, trained on phoneme consonant and vowel units and their fusion respectively. Our experiments show that the deep learned consonant-based acoustic characteristics lead to better recognition results than vowel-based ones. The fusion of vowel and consonant speech characteristics through a deep network significantly outperforms the single space networks as well as the state-of-art deep learning approaches on the DAIC-WOZ database.

\end{abstract}

\begin{graphicalabstract}
\begin{center}
\fbox{
\begin{tabular}{p{.4\textwidth}p{.5\textwidth}}
\includegraphics[width=.3\textwidth]{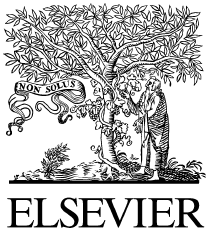}
& 
Depression is a common and serious mood disorder that negatively affects the patient's capacity of functioning normally in daily tasks.
Speech is proven to be a vigorous tool in depression diagnosis. 
Research in psychiatry concentrated on performing fine-grained analysis on word-level speech components contributing  to the manifestation of depression in speech and revealed significant variations at the phoneme-level in depressed speech. On the other hand, research in Machine Learning-based automatic recognition of depression from speech focused on the exploration of various acoustic features for the  detection of depression and its severity level. Few have focused on incorporating phoneme-level speech components in automatic assessment systems. In this paper, we propose an Artificial Intelligence (AI) based application for clinical depression recognition and assessment from speech. We investigate the acoustic characteristics of phoneme units, specifically vowels and consonants for depression recognition via Deep Learning. We present and compare three spectrogram-based Deep Neural Network architectures, trained on phoneme consonant and vowel units and their fusion respectively. Our experiments show that the deep learned consonant-based acoustic characteristics lead to better recognition results than vowel-based ones. The fusion of vowel and consonant speech characteristics through a deep network significantly outperforms the single space networks as well as the state-of-art deep learning approaches on the DAIC-WOZ database.   
\end{tabular}
}
\end{center}

\end{graphicalabstract}

%\begin{graphicalabstract}
%\includegraphics[width=.95\textwidth]{figures/grabs.pdf}
%\end{graphicalabstract}

\begin{highlights}
 \item A new approach for speech-based automatic depression assessment via deep learning.
 \item Investigate acoustic characteristics of speech consonant and vowel spaces and their fusion for depression recognition.
 \item Consonant-based deep learned acoustic descriptors outperforms vowel-based descriptors.
 \item Fusion of consonant and vowel space deep learned descriptors outperform single space features.
 \item Our proposed method outperforms all state of the art approaches in depression recognition on DAIC-WOZ dataset.
\end{highlights}

\begin{keywords}
Major Depressive Disorder \sep Clinical depression Detection \sep AI-based application \sep HCI-based Healthcare \sep Speech Depression Recognition \sep Deep phoneme-level learning
\end{keywords}

\maketitle

\section{Introduction}
\label{sec1}

Depression (Major Depressive Disorder) is a common and serious mood disorder that negatively affects how a person feels, thinks and acts.  
It leads to a variety of emotional and physical problems, negatively affecting a person's ability to function in daily tasks \citep{kanter2008nature}. 
It is manifested through persistent feelings of sadness, hopelessness and interest in activities loss and considered as a major cause of suicide affecting more than 300 million people  in the world \citep{worlddepression2017}. 
Depression causes a non-negligible economic burden per year due to increasing absenteeism and reduced productivity in the workplace \citep{greenberg2015}. 

If correctly diagnosed, depression is a treatable disorder and its symptoms can be alleviated. 
Yet, misdiagnosing depressed patients is a common barrier \citep{mitchell2009}. 
Mostly, questionnaire based scales are used by clinicians to measure depression \citep{williams2005,williams1988structured,zigmond1983},
Misdiagnosing depression is either due to comorbid conditions presenting similar symptoms such as hypothyroidism, hypoglycemia or even normal stress due to busy daily work, 
or subjective biases due to the clinician's skill as well as the reliability of a patient's perception of their mental state \citep{blais2009understanding}. Automatic depression detection methods can offer an objective framework for depression diagnosis as well as an initial entry point for clinicians.

In the past decade, automatic depression detection has became popular with the emergence of publicly available research datasets \citep{gratch2014} and the success of Machine Learning techniques in learning complex patterns in such applications. State-of-the art approaches proposed to assess depression based on behavioral non-verbal and verbal cues \citep{pampouchidou2017}. Particularly, speech has proven to be a reliable objective marker for depression assessment \citep{cummins2015review} due to its accessibility and availability compared to other modalities. 
For instance, research findings showed that speech segments of depressed subjects are slowed, pauses between two segments are lengthened and differences in timbre were detected \citep{kraepelin1921, pampouchidou2017b} compared to healthy subjects. Furthermore, the rise of deep learning techniques had led researchers to explore different deep architectures and methods for depression assessment \citep{EmnaRejaibi2019, salekin2018}. Deep neural networks have significantly improved the performance of such tasks compared to traditional approaches due to their capacity to automatically learn both low and high level descriptors from the patient's behaviors without the need of human intervention.

A sequence of speech sounds is organized in function of organization units called syllables, which are typically constituted of phoneme units, typically, vowel and consonant units \citep{de2003temporal}.
Speech vowels and consonants duration is directly influenced by factors such as speaking rate, adjacent consonants, syllabic stress, position of a vowel in a word and the word type \citep{farooq2018}. 
Consequently, speech vowels and consonants are directly affected in depressed speech. 

Vowel-level analysis in depressed patients was studied in previous work. Specifically, findings revealed a reduced frequency range in vowel production \citep{darby1984speech} and in the speech Vowel Space Area (VSA) \citep{scherer2015self} in depression. Moreover, gender-dependent vowel level analysis was performed for boosting speech-based depression recognition \citep{vlasenko2017implementing}.
Surprisingly, these studies ignored the speech consonant space. 

In this paper, we focus on investigating the contributions of acoustic characteristics of phoneme units for depression assessment. We explore the capacity of deep learning to abstract high level descriptors from speech consonant and vowel spaces.  We propose and compare three spectrogram-based deep Neural Network architectures, trained on speech consonant space, vowel space and deep consonant-vowel fusion respectively. 
The structure of the paper is organized as follows: the next section illustrates the related work. Section~\ref{sec3} represents the the global schema and the proposed CNN architectures. In Section~\ref{sec4}, details about different experiments and results are given. Section~\ref{sec5} concludes the paper.

\section{Related Work}
\label{sec:related_work}
Traditionally, interview style questionnaire based scales are used by health practitioners to assess depression. 
Example scales include: Hospital Anxiety and Depression Scale (HADS) \citep{zigmond1983}, Hamilton Depression Rating Scale (HDRS) \citep{williams1988structured}, 
Quick Inventory of Depression Symptomatology (QIDS) \citep{rush2003}, Beck's Depression Inventory (BDI) \citep{olaya2010}, and the most commonly used, Patient Health Questionnaire (PHQ) \citep{williams2005}. PHQ is composed of nine clinical questions. Based on the patient answers, PHQ score is used to describe the severity level ranges on a scale of $0$ to $23$. 
Patient Health Questionnaire (PHQ) \citep{williams2005} is the most commonly used scale. PHQ is composed of nine clinical questions. Based on the patient answers, PHQ score is used to describe the severity level ranges of depression on a scale of $0$ to $23$.
% commented for now to save space, can be added in the revised version.
 
The assessment rely on clinicians subjective assessments which might  present subjective biases \citep{blais2009understanding} due to the clinician's skill and the reliability of a patient's perception of their mental state \citep{blais2009understanding}.
Consequently automatic depression assessment approaches emerged which offer an objective way of mapping patient's verbal and non-verbal cues to a depression score.

Automatic depression assessment approaches are constituted mainly of three steps: 1) \textit{data collection} where multi-modal data is acquired (audio, video \citep{deencoding,song2020spectral}, text \citep{al2018detecting}, context \citep{ware2020predicting}, etc..) and  depression score ground truth is collected simultaneously using clinically validated scales. 2) \textit{data processing}  where data is pre-processed, depression  markers extracted (features) and 3) \textit{prediction} where machine learning models are applied to predict the individual's depression state.

Speech-based approaches for depression assessment rely on the extraction of acoustic and prosodic markers. 
These include low and high level features designed and extracted from the audio signal in an attempt to model the characteristics of speech such as prosody, voice quality, frequency range, energy, etc. 
Extracted features are then fed to main-stream classifiers to predict depression \citep{alghowinem2013, jiang2018, liu2015, ringeval2017, valstar2016}. 
%Several features were proposed in the literature for detecting depression \cite{jiang2018,low2010,vignolo2016}. 

\subsection{Acoustic features based depression assessment}
Acoustic features used for depression assessment can be categorized into six categories: Prosodic, Source, Formant, Spectral  \citep{cummins2015review}, Cepstral and deep learning features. These features have been demonstrated to contain relevant information about depressed speech.

\textbf{Prosodic features} represent phoneme level variations in speech rate, rhythm, loudness, intonation and stress \citep{al2018detecting, asgari2014,jiang2018,ringeval2017}. For instance, the Fundamental frequency (F0) and energy is used as it represents pitch and loudness perceptual characteristics \citep{cummins2015review}.

\textbf{Source features} capture information of the voice production source. Such features parameterize the air flow from the lungs through glottis via glottal features \citep{jiang2018,low2010}, or vocal fold movements via voice quality features. These include: spectrum maxima, normalized amplitude quotient, the quasi open quotient, harmonic differences H1-H2 and H1-H3, jitter, and the shimmer \citep{asgari2014,ringeval2017,valstar2016}. Another type of source features is the Teager Energy Operator (TEO) \citep{low2010} which is used for computing the audio signal's energy in nonlinear manner \citep{sundaram2003instantaneous}. `TEO measures the number of additional harmonics due to the nonlinear air flow in the vocal tract that produces the speech signal' \citep{low2010}. The interest of this filter lies in the fact that it has a small time window making it ideal for signals local time analysis \citep{kvedalen2003signal}.

\textbf{Formant features} contain information concerning the vocal tract acoustic resonances and articulatory efforts. These reflect physical vocal tract properties such as the muscle tension in the form of formant frequencies (F1, F2, F3) that are affected by the depression state of the patient \citep{al2018detecting,cummins2017}.

\textbf{Spectral features} characterise the speech spectrum which constitutes frequency distribution of the speech signal at a specific time instance \citep{al2018detecting,asgari2014,jiang2018,ringeval2017}.
Examples of spectral features used in the literature include spectral flux, energy, slope and flatness  \citep{jiang2018,low2010, song2018}. 

\textbf{Cepstral features} are those based on a non-linear spectrum-of-a-spectrum representation. The most common used are Mel-Frequency Cepstral Coefficients (MFCC) \citep{asgari2014} and Linear Prediction Cepstral Coefficients \citep{alghowinem2013, jiang2018, lopez2015, low2010}. 
Many authors  reported that spectral features attained a more robust performance for depression detection compared to cepstral and other acoustic features \citep{france2000, low2010, moore2004, moore2007}.

\textbf{Deep Learning features} offer an automatic abstraction of audio descriptors contributing to the manifestation of depression in speech, which has proven to outperform others for depression recognition.
Raw audio is either used as input to deep architectures \citep{trigeorgis2016} or acoustic features are extracted a priori and used as input of the deep  neural network. %For example, \cite{trigeorgis2016} used the raw audio data for the deep learning neural network. 
For instance, \cite{dinkel2019depa} proposed a self-supervised, Word2Vec like pre-trained depression audio embedding method as a feature for depression detection.  
%\textcolor{cyan}{Added this because we compare to them}
On the other hand, \cite{ma2016} used Mel-scale filter bank and MFCC features while in \cite{EmnaRejaibi2019}, a spectogram-based CNN and an MFCC-based CNN are fused to detect depression.
\cite{salekin2018} extracted prosodic and MFCC features, \cite{rejaibi2019} extracted MFCC features and \cite{yang2017,yang2017b} performed textual content analysis, visual features,  spectral and voice quality audio features a priori to Deep Learning.

\textbf{Deep architectures -- }%Several deep learning architectures have been proposed in the literature. 
Proposed deep learning architectures  include feed-forward neural network (FF-NN) \citep{dham2017}, convolutional neural network (CNN) \citep{EmnaRejaibi2019}, long short-term memory convolutional neural network (LSTM-CNN) \citep{ma2016, trigeorgis2016},  bidirectional long short-term memory convolutional neural network (BLSTM-CNN) \citep{dinkel2019depa,salekin2018}, deconvolutional neural network (DNN) \citep{gupta2017, kang2017}, long short-term memory recurrent neural network (LSTM-RNN) \citep{al2018detecting, rejaibi2019}, BLSTM-RNN \citep{salekin2018}, deep convolutional neural network- deconvolutional neural network (DCNN-DNN) \citep{yang2017} and deconvolutional neural network multiple instance learning (DNN-MIL) \citep{salekin2018}. 
To the best of our knowledge the DCNN-DNN model proposed by \citep{yang2017} and LSTM-RNN model proposed by \citep{ rejaibi2019} outperformed all the existing approaches.

\subsection{Phoneme-level Analysis} 
Most of the above approaches extract acoustic features from the entire speech acoustic signal. Few have concentrated on performing fine-grained analysis on the speech components contributing the most in the manifestation of depression in speech and incorporating that in automatic assessment systems. 
Research in psychiatry concentrated on underpinning how speech is affected in depressed patients. \cite{flint1993abnormal} studied the difference of speech through an analysis of phoneme-level variations between healthy subjects and depressed patients. The study revealed significantly shortened voice onset time (consonant/vowel time length) and decreased second formant transition (frequency variation of vocal tract shape during production of a diphthong) and an increased spirantization (presence of aperiodic waves, not attributable to background noise, during the closure interval of stop consonants) compared to controls. 

\cite{darby1984speech} reported a reduction in frequency range of vowel production in individuals suffering from neurological and psychological disorders. \cite{esposito2016significance} concentrated on pauses and spectrogram-based consonant/vowel lengthening analysis to investigate how speech is affected in depression.
Building on psychiatric research, \cite{cummins2017,vlasenko2017implementing} focused on vowel-level analysis to investigate the effect of gender differences in depression and incorporated their findings in vowel-based gender-specific automatic depression recognition system. The authors ignored the speech consonant space in their analysis.
Similarly, in the field of emotion recognition, \cite{ringeval2008exploiting} build KNN classifiers based on consonant and vowels phoneme units features characterization.  
In this paper we investigate the reliability of phoneme-level features extracted from consonant and vowel spaces as objective markers of depression. We extend the state of art in proposing three spectrogram-based deep CNN architectures built on consonant, vowel and consonant/vowel spaces fusion.

\section{Motivations and Contributions}
%features, spectral are the best :P
A variety of acoustic features (prosodic, source, formant, spectral and cepstral) can be extracted from the audio signal, as shown in the related work (section \ref{sec:related_work}), to characterize the speech signal for depression recognition. Spectral features, offering a time-frequency representation of the audio signal, attained robust performance for this task as compared to other audio features \citep{EmnaRejaibi2019}.

% classification: deep is the best
To predict depression, acoustic features are typically fed to shallow machine learning models (eg. Support Vector Machines (SVM), Gaussian Mixture Models (GMM), etc.) \citep{alghowinem2013} or deep learning models. Compared to shallow models, deep learning models have shown higher performances in the recognition of depression due to their capacity to automatically learn both low and high level descriptors from the patient's behaviors without the need of human intervention.

Research in psychiatry concentrated on performing fine-grained analysis on word-level speech components contributing to the manifestation of depression in speech and revealed significant variations at the phoneme-level. On the other hand, research in Machine Learning-based automatic recognition of depression from speech focused on the exploration of various acoustic features for the  detection of depression and its severity level. 
While some features such as prosodic features represent phoneme-level variations of the speech signal, most of the state-of-art approaches extract acoustic features from the entire speech  signal, and 
few  have focused on incorporating phoneme-level speech components in automatic depression assessment systems. Other than \citep{vlasenko2017implementing} which proposed a vowel level analysis framework for speech-based depression recognition, it has not been reported in the literature any work exploring  phoneme-level units (consonants and vowels) as objective markers in a deep learning framework for depression recognition.

In this work, the reliability of phoneme-level features extracted from consonant and vowel spaces as objective markers of depression is investigated. Due to their reported efficiency, spectrogram-based features are extracted from the consonant and vowel spaces before feeding them into 2 deep architectures and fusing them for depression classification. 
Compared to the state-of-the-art, the presented study is the first to explore phoneme-level units in a deep learning framework for depression recognition.
The contributions of this work can be summarized by the following.
\begin{itemize}
\item A new high-performing phoneme-level based approach for speech-based automatic depression assessment via deep learning.

 \item Investigation of the acoustic characteristics of speech consonant and vowel spaces and their fusion for depression recognition.

 \item Studying and comparing the performances of the consonant-based and vowel-based deep learned acoustic descriptors for depression assessment.
 
 \item Investigating the performances of the fusion of consonant and vowel spaces deep learned descriptors compared to single space features.
 
\end{itemize}

\begin{figure}[]
\centering
\includegraphics[width=1\linewidth]{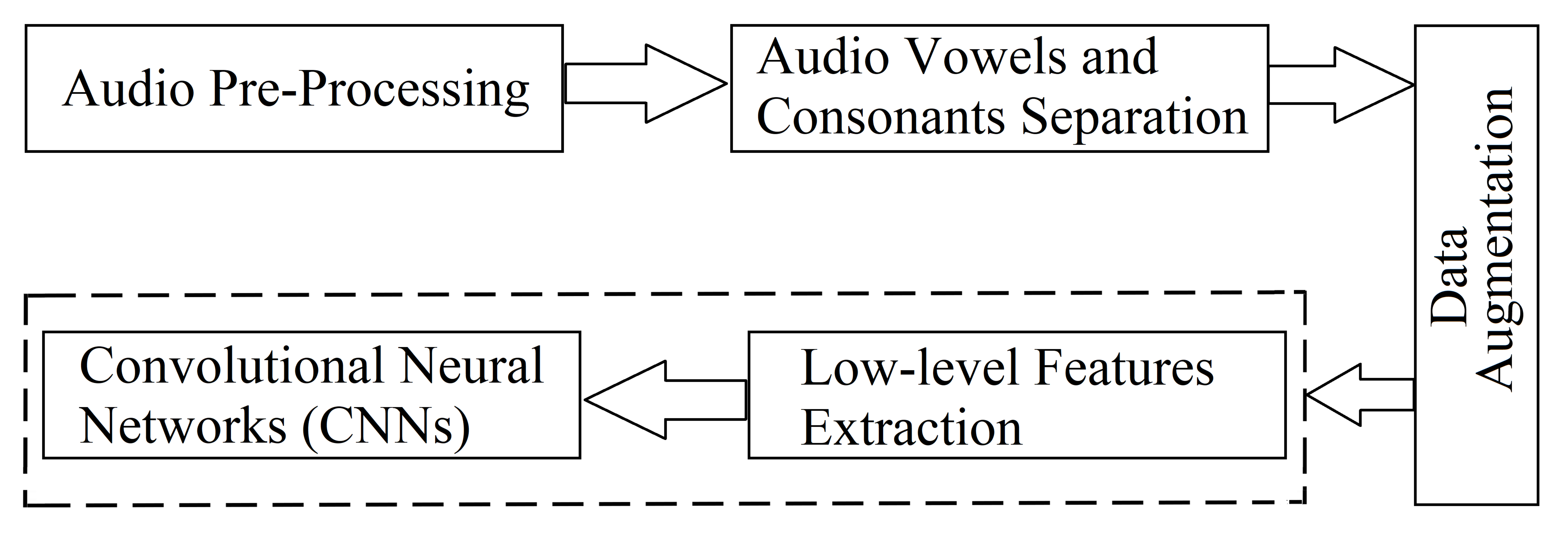}
\caption{Block diagram of proposed methodology for audio vowels and consonants based depression assessment.} %First, the audio signal is pre-processed and the patient's speech is extracted. Then, the vowels.} 
\label{fig:basic}
\end{figure}

\section{Proposed Method}
\label{sec3}
We propose a novel approach that automatically learns and fuses vowel and consonant spaces acoustic features via a deep learning architecture.  
The proposed approach is constituted of four steps as shown in Figure \ref{fig:basic}.

First, patient's speech is extracted from the sound signal in a pre-processing step (Section~\ref{sec:preprocessing}). Second, voiced segments are extracted from which vowels and consonants separation  is performed (Section~\ref{sec:audio_vowels_and_consonants_separation}). Then, data augmentation is performed to overcome over-fitting issue relevant to training Deep Learning Networks (Section~\ref{sec:data_augmentation}). Finally, acoustic characteristics of speech vowel and consonant spaces are extracted and fused through a deep Convolutional Neural Network (CNN) architecture, AudVowConsNet, for depression detection and severity level recognition.  (Section~\ref{sec:AudVowelConsNet_Architecture}).

\subsection{Patient Speech Extraction}
\label{sec:preprocessing}

Our approach is based on audio recordings extracted from clinical interviews which constitute conversations between a health practitioner and a patient responding to a validated questionnaire intended to measure depression level.
Our approach is based on depression assessment from the patient's responses to the clinical questions.
Thus, we pre-process the recordings to extract the patient's speech from that of the interviewer.

\subsection{Audio Vowels and Consonants Separation}
\label{sec:audio_vowels_and_consonants_separation}

\begin{figure}[]
\centering
\includegraphics[width=0.8\linewidth]{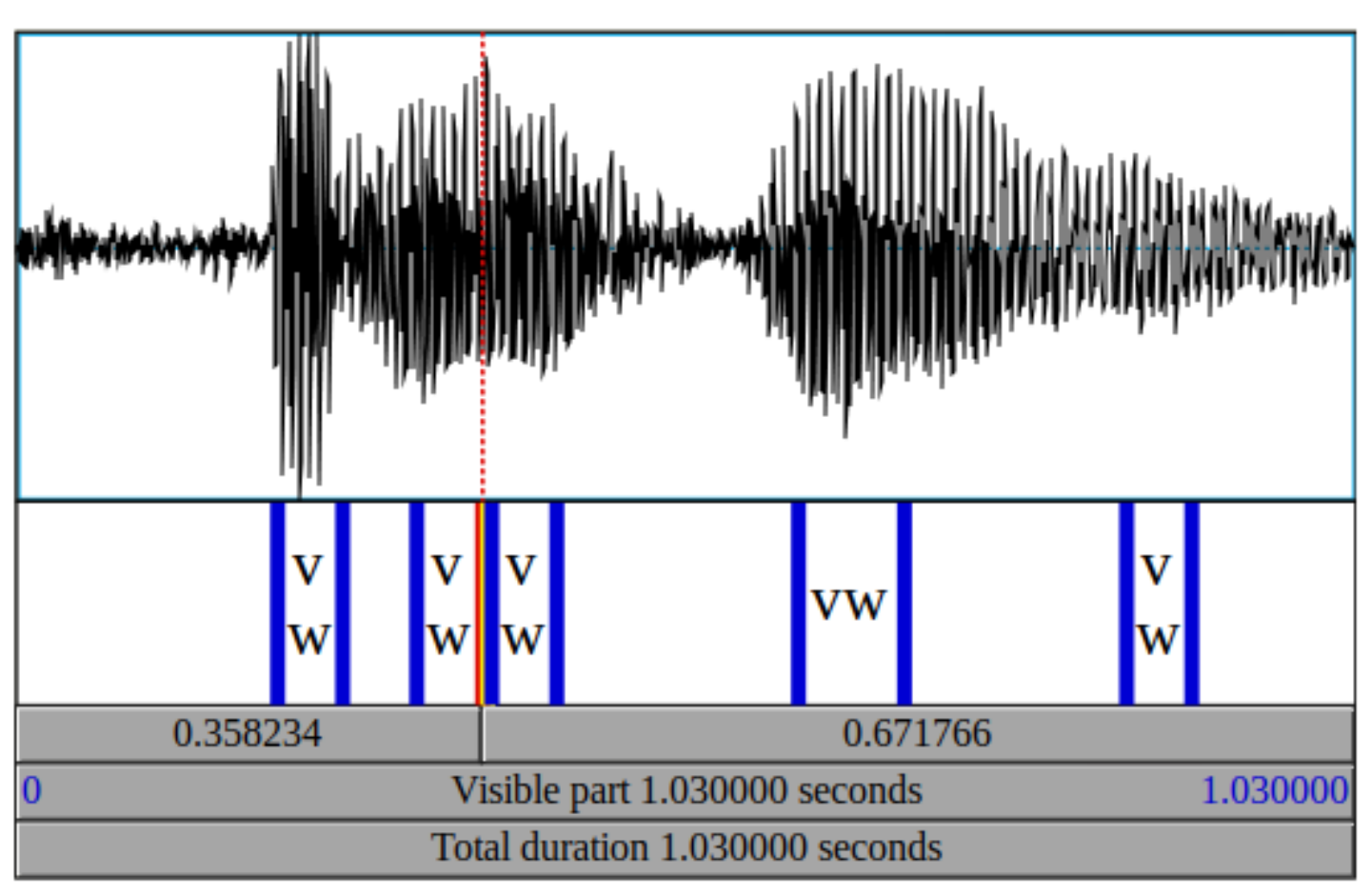}
\caption{Separation of audio vowels and consonants using Praat software. Dark blue lines shows the boundaries of vowels.}
\label{fig:vowel}
\end{figure}

Patient's vowels and consonants spaces are extracted through two steps. First voiced and unvoiced segments are separated from the patient's speech. Then, vowels and consonants are extracted from voiced sounds based on phonetic transcription alignment. 
\begin{enumerate}
    \item \textbf{Speech Segmentation --} 
Voiced and unvoiced frames are separated using a threshold-based method on time frames of $0.01$ second in Praat software \citep{boersma2001praat}. The following separation threshold values are selected based on previous literature studies \citep{boersma2001praat}:

\begin{itemize}
\item %Pitch floor is set to $75 Hz$ and the pitch ceiling is set to $500 Hz$. 
Frames with pitch higher than a pitch ceiling of $500 Hz$ are considered as voiceless. %to adjust the window size. %\textcolor{cyan}{what window size? you mean frame size?}\textcolor{magenta}{Window size and frame size are two different parameters. }
\item %A Silence Threshold (ST) is computed for every frame. 
Frames that do not contain signal amplitudes above a Silence Threshold (ST) are considered voiceless. 
Let $PH$ be the amplitude of the highest frame peak  and $GP$ the amplitude of highest peak in the patient's audio file, then ST is given by: $ST= HP -{{0.03}\times{GP}}$

\item %A Voicing Threshold (VT) is set to $0.45$. 
A frame is marked as unvoiced if the signal strength of all its samples do not exceed a Voicing Threshold (VT) of $0.45$. 
\item  To account for consecutive voiced or unvoiced segments, a Voiced/Unvoiced Cost (VUC) is set to $0.14$. 
\end{itemize}

\item \textbf{Vowels and Consonants Separation} -- %Voiceless segments of the participant are discarded and voiced segments are further divided into vowels and consonants. 
Vowels and consonants are extracted from voiced segments based on phoneme level transcription alignment. This requires a corresponding lexicon containing phonetic transcription of words present in the corpus. %Audio files does not provide such a lexicon. 
We use the Carnegie Mellon University (CMU) pronouncing dictionary \citep{lenzo2007} transcriptions. The phoneme alignment and vowel space plotting were performed using Praat vocal toolkit and R vowels package \citep{praat2019}, (cf. Figure \ref{fig:vowel}).   
\end{enumerate}

\subsection{Data augmentation}
\label{sec:data_augmentation}
Data augmentation is performed to overcome the problem of over-fitting and data scarcity and to improve the proposed architecture robustness against noise. 
Two types of audio augmentation techniques are considered to perturb the consonant and vowel audio signals: Noise Injection and Pitch Augmentation \citep{rejaibi2019}. 
\begin{itemize}
    \item \textbf{Noise Injection}: random noise is added to the audio signals. If y is the audio signal and $\alpha$ is the noise factor, then the noise augmented data $x$ is given by:
$x=y-{{\alpha}\times{rand(y)}}$. We use $\alpha$ = $0.01$, $0.02$ and $0.03$.
\item \textbf{Pitch Augmentation}: Audio sample pitch is lowered by $0.5$, $2$, $2.5$ (in semitones).%different amounts.%(duration kept unchanged).
\end{itemize}

\begin{figure*}[pos=H,width=1\textwidth,align=\centering]%[H]
\centering
\includegraphics[width=\textwidth]{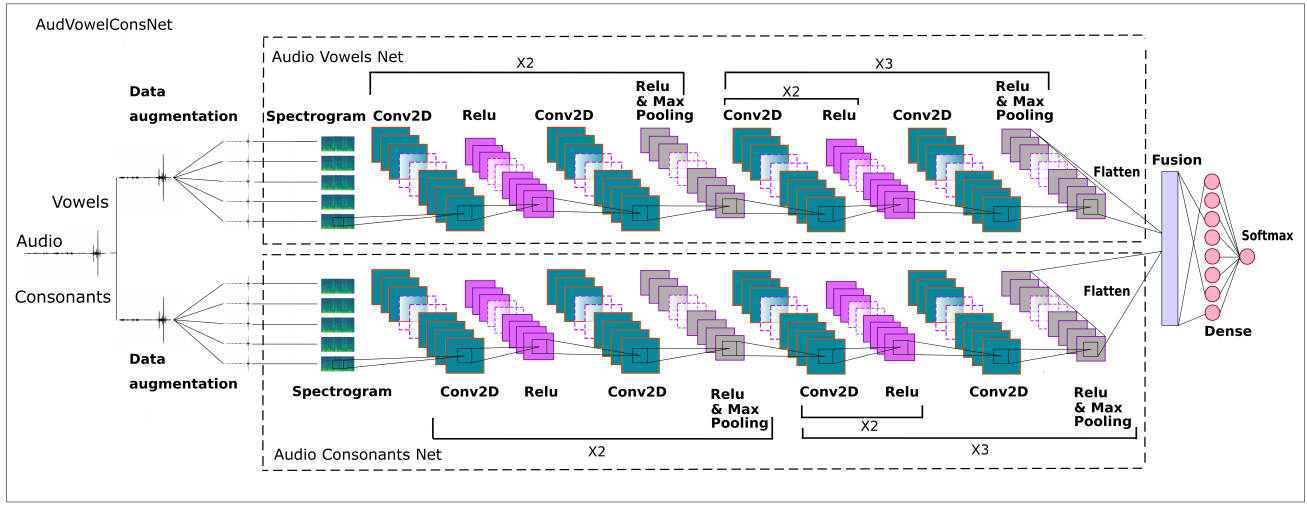}
\caption{The proposed AudVowelConsNet architecture: a phoneme-level based CNN that fuses vowel and consonant acoustic features via a deep learning framework.} 
\label{fig:AudVowelConsNet_Architecture}
\end{figure*}

\subsection{AudVowConsNet Architecture}
\label{sec:AudVowelConsNet_Architecture}
The proposed deep neural network named as AudVowConsNet is the fusion of the vowels and consonants spectrogram-based CNNs, as shown in Figure \ref{fig:AudVowelConsNet_Architecture}. The output from the both architectures are flattened and fused in a fully-connected layer in order to generate the label prediction for binary (0 for non-depression and 1 for depression) and $24$ depression severity levels.

\subsubsection{Spectrogram-Based Low-level Features Extraction}
\label{sec:low_level_features_extraction}
Spectral features attained robust performance for depression detection as compared to other audio features \citep{EmnaRejaibi2019}. 
To extract spectrogram from the  original and augmented audio files, Short-Time Fourier Transform (STFT) is applied to the audio data. Let, $x[n]$ is the audio signal, then the STFT f(n,$\omega$) is given by:
\begin{equation}
f(n,\omega)=\sum_{n=-\infty}^{\infty}{x[n]}{w[n-m]}e^{-j{\omega}m}
\end{equation}
f(n,$\omega$) is the output of STFT and $w[n]$ is the sliding window that emphasizes the local frequency parameter. The Hann window is selected as the sliding window and its equation is given bellow:
\begin{equation}
w(n)=0.5\left(1-cos\left(2\pi\frac{n} {N-1}\right)\right)
\end{equation}
Where N represents the length of the window and 0 $\leq$ n $\leq$ N.
The spectrogram is computed by squaring the magnitude of the STFT signal f(n,$\omega$):
\begin{equation}
S(n,\omega)= |f(n,\omega)|^2
\end{equation}

The obtained spectrograms are fed to the deep convolution neural networks to extract high level features.

\subsubsection{Deep learned High-level Features Descriptors}
\label{sec:high_level_features_extraction}
The low-level spectral features extracted from the audio vowels and consonants are fed to two different Convolutional Neural Networks (named: Audio Vowels Net and Audio Consonants Net) to extract high-level features as shown in Figure ~\ref{fig:AudVowelConsNet_Architecture}. Each of these modalities consist of five convolutional blocks with same layout structure. First two blocks composed of a two-dimensional (2D) convolutional layer followed by a ReLU activation function, a second convolutional layer, a ReLU, and maxpooling layer. While, the remaining three blocks consist of convolutional layer, ReLU activation function, a second convolutional layer, a ReLU, a third convolutional layer, another ReLU, and finally a maxpooling layer. The feature vector size of individual network is 8194.

\section{Results and Discussion}
\label{sec4}

In this section, we present the different experiments we performed for depression assessment from speech signals using the proposed deep architecture. 
We compare the performance of $3$ models: 1) Audio Vowels Net trained on the audio vowel space, 2) Audio Consonants Net trained on audio consonant space and 3) AudVowelConsNet Net, trained on the fusion of Audio Vowels Net and Audio Consonants Net.
%Moreover, we compare the performance of the three networks with different state-of-the art approaches.
Moreover, we compare the performance of the three proposed networks with state-of-the art benchmark approaches.
In the following we present the details concerning the performed experiments.
%\begin{enumerate}
%    \item AudVowNet: Vowels-based CNN architecture
%    \item AudConsNet: Consonants-based CNN architecture
%    \item AudVowConsNet Net: Fusion of Audio Vowels and Audio Consonants Net.
%\end{enumerate}

\begin{table*}[pos=H,width=0.87\textwidth,align=\centering]
\centering
\caption{\label{tab_test_res} Proposed deep neural networks (Audio Vowels Net, Audio Consonants Net and AudVowConsNet) performances for depression assessment tasks (PHQ-8 Binary and Score) in terms of Accuracy, RMSE, CC, and CCC.}
\begin{tabular}{|l||c|c|c||c|c|c|}
\hline
\textbf{Network}  & \multicolumn{3}{c||}{\textbf{PHQ-8 Binary}} & \multicolumn{3}{c|}{\textbf{PHQ-8 Score}}\\
\cline{2-7}
 & Accuracy (\%) & CC & CCC & Accuracy (\%) & CC & CCC\\
\hline \hline
Audio Vowels Net & 78.77 & 0.58& 0.57 & 54.26  & 0.59& 0.58\\%0.1690 vs 0.17395942
Audio Consonants Net& 80.98&  0.62&  0.61 & 57.57 & 0.61  & 0.61\\
AudVowelConsNet   & \textbf{86.06} & \textbf{0.72} & \textbf{0.72}& \textbf{70.86} & \textbf{0.73} & \textbf{0.73}\\
%\cline{2-10}
\hline
\end{tabular}
\end{table*}

\subsection{Dataset}
\label{sec:dataset}
To evaluate our proposed model, we use the Distress Analysis Interview Corpus Wizard-of-Oz dataset (DAIC-WOZ) \citep{gratch2014}. 
%DAIC-WOZ was introduced in Audio/Visual Emotion Challenge and Workshop, 2017 (AVEC 2017) \cite{gratch2014}. 
%It is collected by University of California and is a part of large DAIC corpus. 
DAIC-WOZ contains clinical interviews recorded to  investigate different psychological distress conditions such as depression, anxiety, and post-traumatic stress disorder. 
The dataset is composed of $189$ audio recordings of participants interviews with a virtual interviewer. All audio recordings are labeled by the PHQ-8 scores (depression severity level, [0-23] range) and the PHQ-8 binary (1/0 depressed vs. not depressed).
The average length of the audio recordings is $15$ minutes with a sampling rate of $16 kHz$. For technical reasons, in this study, only $182$ participants audio recordings are used. 

For all experiments, the dataset is randomly divided into $80\%$ for training, $10\%$ for validation and $10\%$ for testing.

\subsection{Evaluation Metrics}
\label{sec:matric}
The proposed approach is evaluated using Precision, Recall, F1-score, Root Mean Square Error (RMSE), Pearson Correlation Coefficient (CC) and Concordance Coefficient Correlation (CCC) \citep{lawrence1989}. 
$CCC$ is a measure of the agreement between the predicted and true depression scores. Let $True$ and $Pred$ be the true and predicted depression score vectors, then $CCC$  is given by:

\begin{equation}
CCC(True,Pred)=\frac{2{CC}{\sigma_{True}}{\sigma_{Pred}}} {\sigma^2_{True}+\sigma^2_{Pred}+(\mu_{True}-\mu_{Pred})^2}
\end{equation}
$\sigma_{True}$ and $\sigma_{Pred}$ represent the standard deviations of variables  $True$ and $Pred$, and $\mu_{True}$ and $\mu_{Pred}$ represents their respective means. 
\subsection{Network Implementation Details}
Both vowels and consonants-based CNN architectures are constituted of five blocks. 
The number of convolution filters ($N$) for each block, from input to output, is set to $N = [64, 128, 256, 512, 512]$. For all convolutional layers, we use a filter size of $3 \times 3$. The stride and  pool size of the max pooling layer are $2 \times 2$.

In both networks, RELU is used as activation function for all convolutional and  fully connected layers. For AudVowelConsNet, two fully connected layers are used with RELU as activation function.    
For all three networks, to predict the PHQ-8 binary, the output layer is a dense layer of size $2$  with a Softmax activation function. 
To predict the PHQ-8 scores (24 levels), an output dense layer of size $24$ neurons is used with a Softmax activation function. 
The proposed models are trained with the RMSE as loss function and the Adam optimizer with a learning rate of $10^-5$ and a decay of $10^-6$. The batch size is set to $120$ samples. The number of epochs for training is set to $500$. An early stopping is performed when the loss function stops improving after $10$ epochs. 

%===============================================================

\subsection{Performance Analysis of Proposed Networks}

All three spectrogram based CNN networks (Audio Vowels Net, Audio Consonants Net and AudVowelConsNet) are trained to deliver the PHQ-8 binary and PHQ-8 scores.  %We report results per audio sample
%The performance of the network is evaluated for both tasks as follows:

\subsubsection{Comparison between the 3 proposed networks}

%%===============Table 2=======================

% on depressed class its higher
\begin{table*}[]
\caption{\label{tab_PHQ-8_binary_D_ND} Comparison of the proposed deep neural networks architectures for PHQ-8 binary assessment for the two binary classes of Depression (D) and Non-Depression (ND).}
\centering
%\begin{tabular}{|p{3.4cm}||p{1cm}|p{1cm}|p{1cm}|p{1cm}|p{1cm}|p{1cm}|p{1cm}|p{1.3cm}|p{1cm}|p{0.7cm}|}
\begin{tabular}{|p{4cm}||c|c|c|c|c|c|c|c|c|}
\hline 
\textbf{Method} & \textbf{Accuracy (\%)} & \textbf{AUC Score } &\multicolumn{2}{c|}{ \textbf{Precision (\%)}} & \multicolumn{2}{c|}{\textbf{Recall (\%)}} & \multicolumn{3}{c|}{\textbf{F1-Score (\%)} } \\
\cline{4-10}
& & & D &ND & D &ND & D &ND &Av.\\
\hline \hline 
Audio Vowels Net &  78.77 & 0.75 & 67 & 84 & 66 &85&66 &85 & 78.99  \\%0.7535
Audio Consonants Net & 80.98 & 0.76 & 73 & 84 & 64 &89&68 &86& 80.30  \\%0.7640
AudVowelConsNet &  \textbf{86.06} & \textbf{0.83}  & \textbf{81} &\textbf{88} & \textbf{73} &\textbf{92}
&\textbf{77} & \textbf{90} & \textbf{85.85} \\%0.8253
\hline 
\end{tabular}
\end{table*}

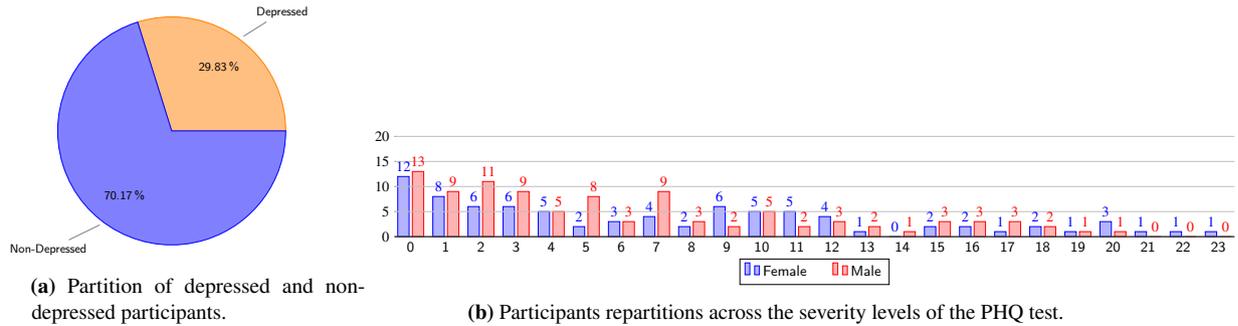
\begin{figure*}
\centering
   \begin{subfigure}[b]{0.25\textwidth}
        %\centering
         \advance\leftskip-0.5cm
         \scalebox{0.5}{\input{./Graphics/PHQ_binary_dispersion}}
         \caption{Partition of depressed and non-depressed participants.}
         \label{figure_rep_1}
   \end{subfigure}
    \begin{subfigure}[b]{0.60\textwidth}
       \scalebox{0.6}{\input{./Graphics/PHQ_scores_dispersion_gender}}
       \caption{Participants repartitions across the severity levels of the PHQ test.}
       \label{figure_rep_2}
    \end{subfigure}
\caption{Depression and severity level repartitions of the participants within the DAIC-WOZ Corpus. (a) Depressed versus Non-Depressed participants. The PHQ-8 binary of Depressed participants is 1 and the PHQ-8 binary of Non-Depressed participants is 0. (b) Participants repartitions across the 24 depression severity levels given by the PHQ-8 test. }    
\label{dataset_repartition}
\end{figure*}

Table \ref{tab_test_res} summarizes the resulting performances of all three networks on the testing set.
% comparison vowels and consonants nets
Results show that Audio Consonants Net performs better in terms of accuracy compared to Audio Vowels Net for both depression assessment tasks: PHQ-8 Binary ($80.98\%$ vs. $78.77\%$) and PHQ-Score ($57.57\%$ vs. $54.76\%$). The higher CC \& CCC confirm the performance. 
%\textcolor{cyan}{Add a reason.}
The improved performance by the consonants net over the vowels one might be due to the fact that consonants require more precise articulation than vowels. Therefore, this  leads to a stronger manifestation of pronunciation differences of consonants in depressed and non-depressed scenarios leading to better results as compared to vowels.

%Compare the fusion 
Fusion of audio vowels and consonants spectrogram-based deep learning descriptors, significantly boosts the performance. 
AudVowConstNet achieves significantly higher accuracy and better CC and CCC compared to both Audio Vowels Net and Audio Consonants Net for both tasks. 
An accuracy of $86.06\%$ and $70.86\%$ are obtained for PHQ-8 binary and PhQ-8 scores, respectively. 
For PHQ-8 binary, fusion model accuracy is improved by $7.99\%$ with respect to Audio Vowels Net and by $5.08\%$ with respect to Audio Consonants Net accuracy. 
Similarly, for PHQ-8 Score, the accuracy is improved by $16.6\%$ and $13.29\%$  with respect to the other two models. 
The Audio Vowels and Audio Consonants Nets separately are not very robust in the assessment of depression. It is the aggregation of both networks that captures  higher-level features which are able to better discriminate the level of depression.
%\textcolor{5}{Figure 5 old place}

The DAIC-WOZ dataset presents an unbalanced number of samples as shown in Figure~\ref{dataset_repartition}. The number of non-depressed participants is three times higher than that of depressed participants. Moreover, the number of samples for each score is not balanced. This imbalance could affect the performance of the proposed models in detecting depression. For that, in Table \ref{tab_PHQ-8_binary_D_ND}, we compare the performance of the PHQ-8 binary proposed networks for Depression (D) and Non-Depression (ND) classes in terms of precision, recall and F-score. We also provide the accuracy and AUC score in the same table. One can notice that AudioVowelConsNet obtained significantly higher values of precision, recall and F1-score for each of the binary classes as compared to both audio vowels and consonants spectrogram based CNN networks. The fusion of consonant and vowel deep learning descriptors significantly enhanced its performance for the classification of Depression and Non-Depression. Furthermore, for the Non-Depression class the model obtained $88\%$, $92\%$ and $90\%$ values of precision, recall and F1 score, respectively. The precision of AudioVowelConsNet for Depression class is higher by $8\%$ and $14\%$ as compared to Audio Consonants Net and Audio Vowels Net, respectively. Moreover, the increment in precision has been observed for Non-Depression class for AudioVowelConsNet as compared to other models.
Furthermore, Audio Consonants Net achieves higher precision for the Depression class as compared to  Audio Vowels Net. For Non-Depression class, the precision value is almost same. 

%\textcolor{red}{Figure 8 new place.}
%%=========================================
\noindent
\renewcommand\arraystretch{1.5}
\setlength\tabcolsep{0pt}
\begin{figure}[!htbp]
\begin{subfigure}[!htbp]{0.4\textwidth}
\centering
\begin{tabular}{c >{\bfseries}r @{\hspace{0.7em}}c @{\hspace{0.4em}}c @{\hspace{0.7em}}l}
  \multirow{10}{*}{\rotatebox{90}{\parbox{1.1cm}{\bfseries\centering Predicted}}} & & \multicolumn{2}{c}{\bfseries Actual} & \\
  & & \bfseries ND & \bfseries D & \bfseries total \\
  & ND & \MyBox{12091}{57.87\%} & \MyBox{2246}{10.75\%} & \MyBoxx{14337}{84.33\% \\15.67\%} \\[2.4em]
  & D & \MyBox{2190}{10.48\%} & \MyBox{4367}{20.90\%} & \MyBoxx{6557}{33.40\% 66.60\%} \\
 & total &14281 & 6613 &\\ 
& & 84.66\% & 33.96\% &\\
&       & 15.34\% & 66.04\% &
\end{tabular}%}\\
\caption{Audio Vowels Net} 
        \label{fig:}
    \end{subfigure} 
\begin{subfigure}[!htbp]{0.4\textwidth}
\centering
\begin{tabular}{c >{\bfseries}r @{\hspace{0.7em}}c @{\hspace{0.4em}}c @{\hspace{0.7em}}l}
  \multirow{10}{*}{\rotatebox{90}{\parbox{1.1cm}{\bfseries\centering Predicted}}} & 
    & \multicolumn{2}{c}{\bfseries Actual} & \\
  & & \bfseries ND & \bfseries D & \bfseries total \\
  & ND & \MyBox{12692}{$60.74\%$} & \MyBox{2386}{$11.42\%$} & \MyBoxx{15078}{$84.18\%$ \\$15.82\%$} \\[2.4em]
  & D & \MyBox{1589}{$7.61\%$} & \MyBox{4227}{$20.23\%$} & \MyBoxx{5816}{$27.32\%$ \\ $72.68\%$} \\
  & total &14281 & 6613 &\\
  & & $88.87\%$ & $36.08\%$ &\\ 
  &       & $11.13\%$ & $63.92\%$ &
\end{tabular}%}\\
\caption{Audio Consonants Net} 
        \label{fig:}
    \end{subfigure} 
\begin{subfigure}[!htbp]{0.4\textwidth}
\centering
\begin{tabular}{c >{\bfseries}r @{\hspace{0.7em}}c @{\hspace{0.4em}}c @{\hspace{0.7em}}l}
  \multirow{10}{*}{\rotatebox{90}{\parbox{1.1cm}{\bfseries\centering Predicted}}} & 
    & \multicolumn{2}{c}{\bfseries Actual} & \\
  & & \bfseries ND & \bfseries D & \bfseries total \\
  & ND & \MyBox{13160}{62.98\%} & \MyBox{1791}{8.57\%} & \MyBoxx{14951}{88.02\% 11.98\%} \\[2.4em]
  & D & \MyBox{1121}{5.37\%} & \MyBox{4822}{23.08\%} & \MyBoxx{5943}{18.86\% \\ 81.14\%}\\
  & total &14281 & 6613 &\\
  & & 92.15\% & 27.08\% &\\
  & & 7.85\% &72.92\% &
\end{tabular}%}
\caption{AudVowelConsNet} 
        \label{fig:}
    \end{subfigure} 
    \caption{Confusion matrix of all three experiments for binary depression assessment tasks.}
\label{fig:confusion}
\end{figure}
%%=========================================

Figure \ref{fig:confusion} shows the confusion matrix for all three experiments in terms of PHQ-binary classes. For AudVowelConsNet a significant increment is observed in true positive (prediction for Depression) and true negative (Non-Depression) values. An increase of $7.49\%$ and $3.28\%$ in true negative predictions (i.e., non-depression class) has been observed as compared to Audio Vowels Net and Audio Consonants Net, respectively. Similarly, for depression class an increment of $6.88\%$ and $9\%$ has been noted as compared Audio Vowels Net and Audio Consonants Net, respectively. Furthermore, a significant decrement in false positives can be seen in  Figure \ref{fig:confusion}. These results show that the fusion features provide a meaningful improvement in the performance of the network.

%\textcolor{red}{Figure 6 old place}

\begin{figure}[]
    \centering
     \begin{subfigure}[t]{.43\textwidth}
        \centering
         \includegraphics[width=1\linewidth]{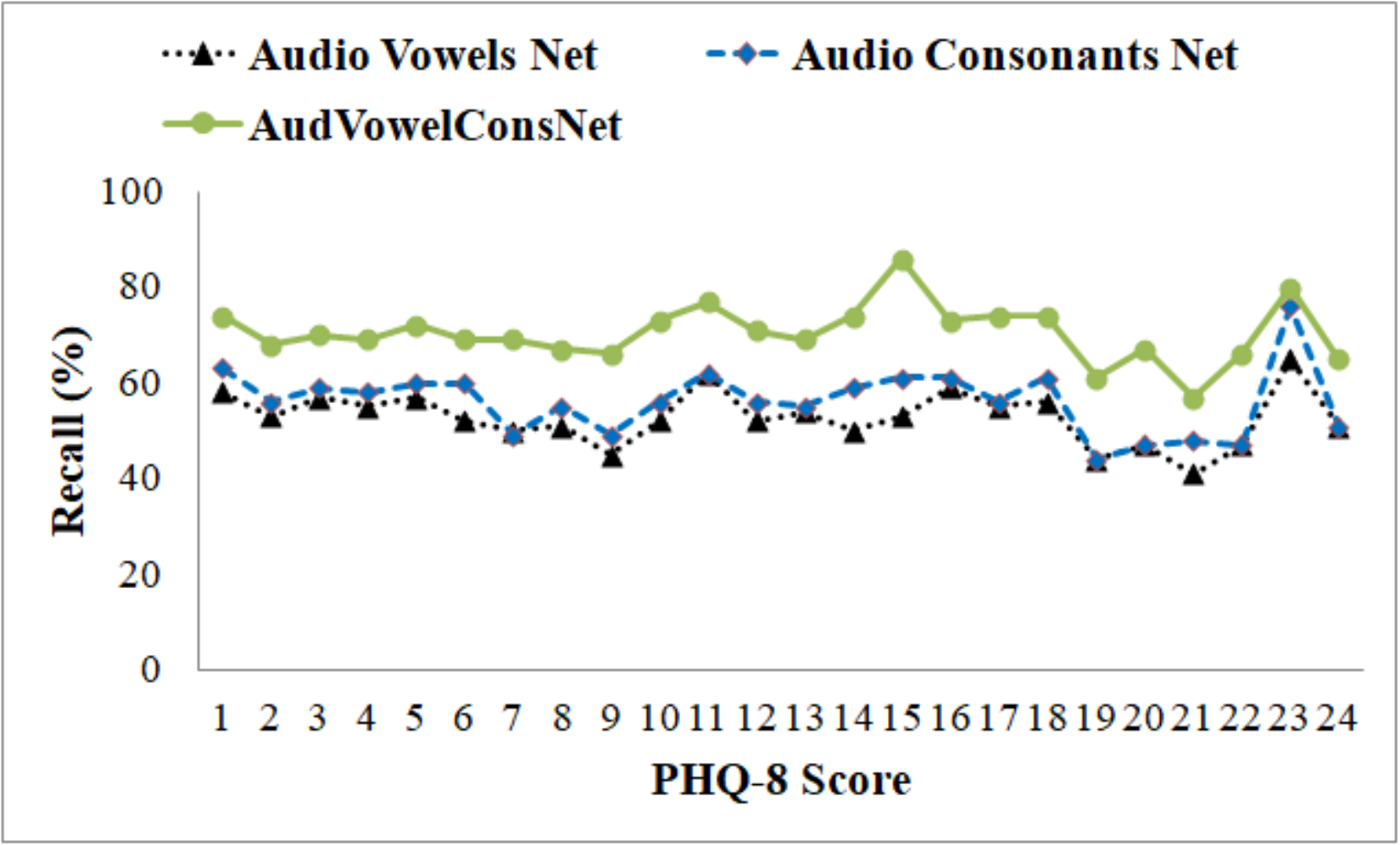}
         \caption{Recall  } 
        \label{fig:PHQ-8_score_Recall}
    \end{subfigure}
    \begin{subfigure}[t]{.43\textwidth}
        \centering
          \includegraphics[width=1\linewidth]{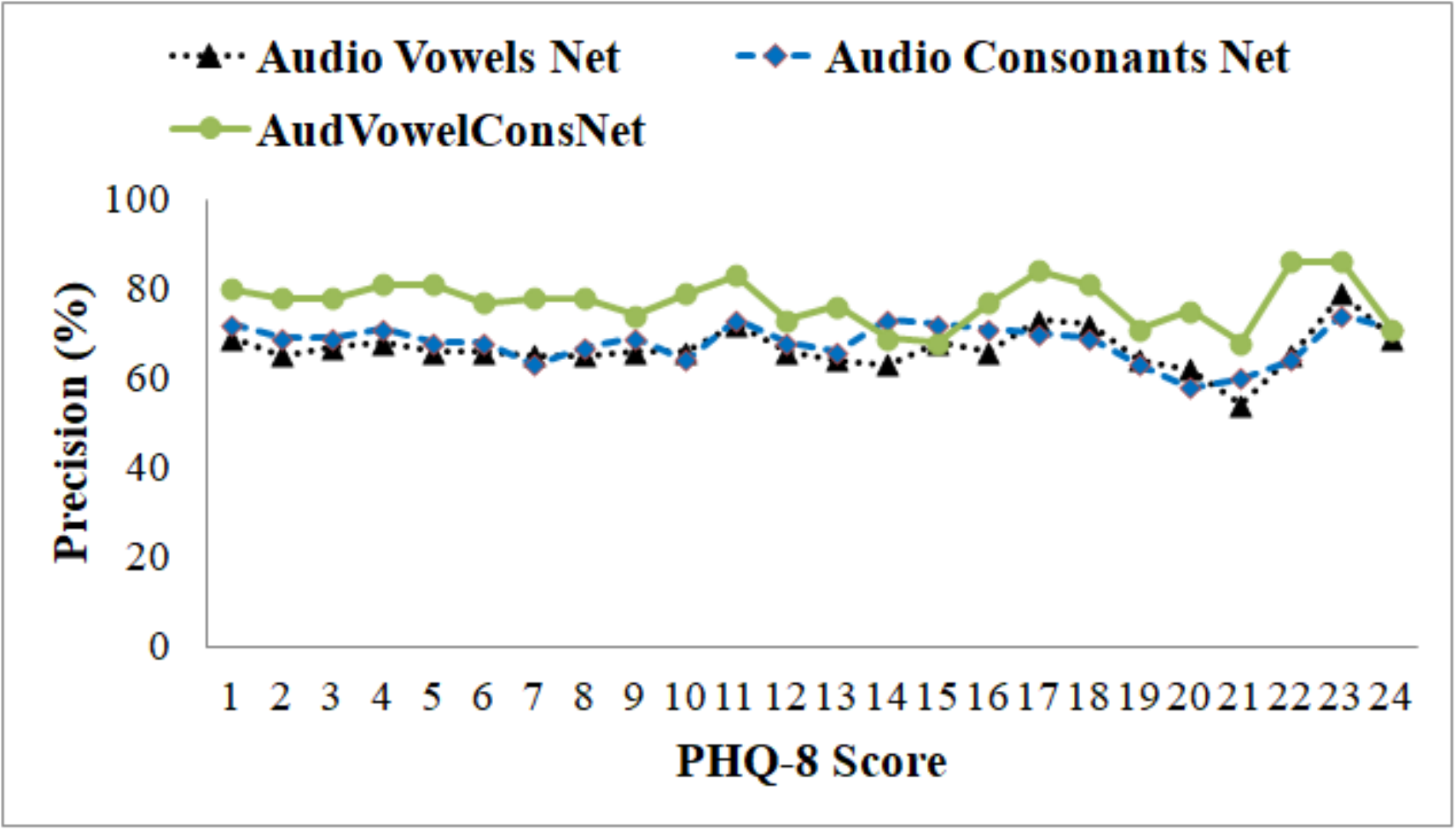}
              \caption{Precision} 
        \label{fig:PHQ-8_score_Precision}
    \end{subfigure}
          \begin{subfigure}[t]{.43\textwidth}
        \centering
          \includegraphics[width=1\linewidth]{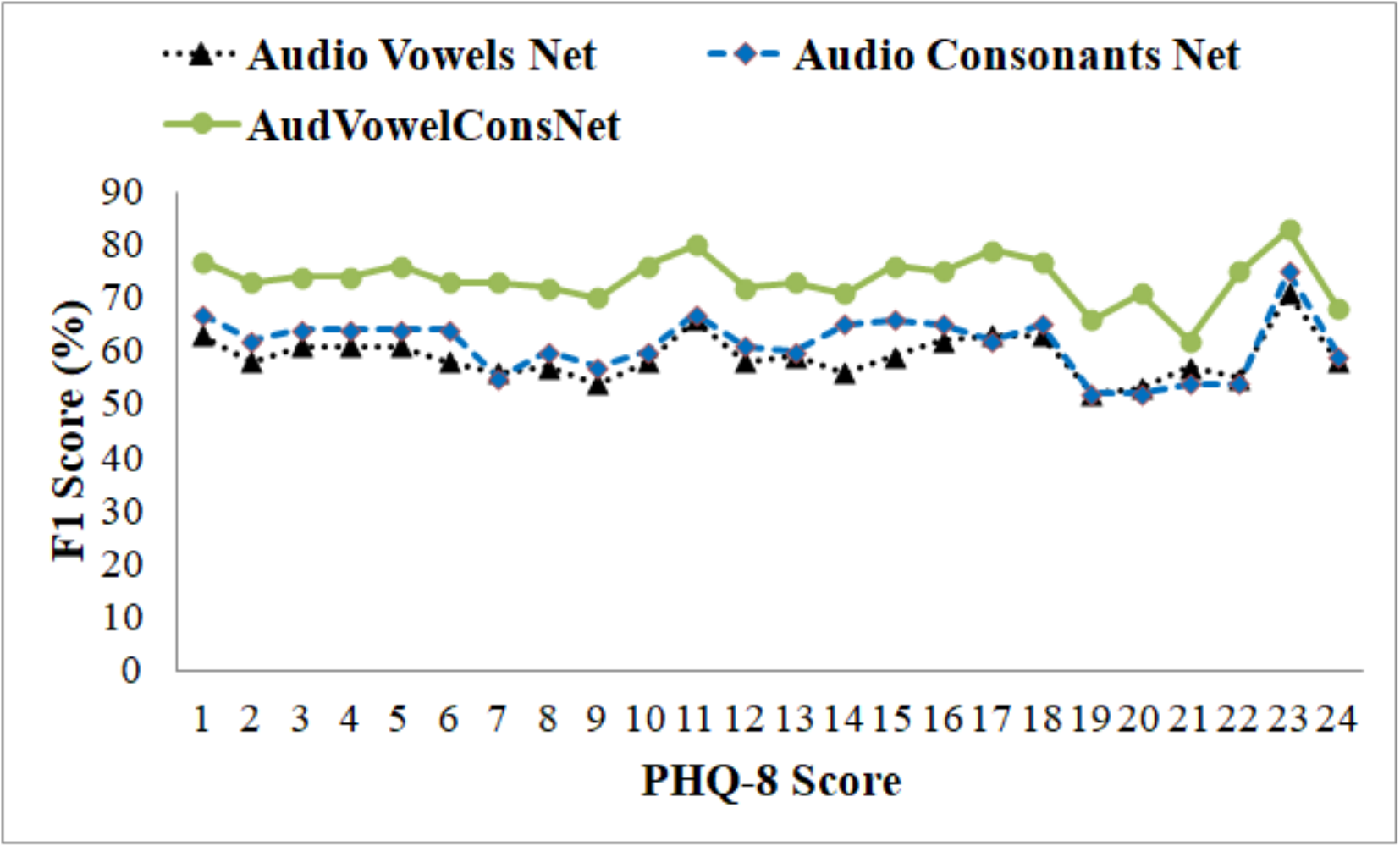}
              \caption{F-score} 
        \label{fig:PHQ-8_score_Fscore}
    \end{subfigure}
\caption{Comparison of the performances of the proposed deep neural networks architectures in the prodiction of the the $24$ PHQ-8 depression severity levels on the test set in terms of Recall, Precision and F-score.
% The weighted mean of the F1-score of the testing samples for each severity level of the three architectures.
}
\label{fig:PHQ-8_score}
\end{figure}

% Compare F-score per severity level
Figure \ref{fig:PHQ-8_score} shows the comparison in terms of F-score, precision and recall between the $24$ PHQ-8 severity levels prediction by the  proposed networks. From the figure, we notice a similar trend in the performance of the three networks. Audio Consonants Net achieves a slightly better performance compared to Audio Vowels Net for all levels. AudVowConsNet outperforms the other networks for all severity levels. Particularly, 
the highest results are obtained for level $23$ by the AudVowConsNet  with an F-score of $83\%$, followed by the Audio Consonants Net ($75\%$) and  Audio Vowels Net ($71\%$).

Regarding Precision, for the AudVowConsNet, the highest precision value of $86\%$ is obtained for severity levels $22$ and $23$ (cf. Figure \ref{fig:PHQ-8_score_Precision}. On the other hand, the lowest precision value of $68\%$ is obtained for levels $15$ and $21$, respectively. For Audio Vowels Net and Audio Consonants Net, the highest obtained precision values are $73\%$ and $74\%$, respectively. Furthermore, the recall values of all three models are compared for each level. The comparison is presented in a Figure \ref{fig:PHQ-8_score}. A notable difference can be observed between the recall values obtained for AudVowConsNet and the values obtained for other models. For AudVowConsNet the highest recall value of $86\%$ is obtained for level $15$. While, the minimum value of $57\%$ is obtained for level $21$.

% compare performance over training ephocs for both

%\textcolor{red}{Figure 5 new place.}
\begin{figure}[]
  \centering
  \begin{subfigure}[t]{0.5\textwidth}
        \centering
        \includegraphics[width=0.75\textwidth]{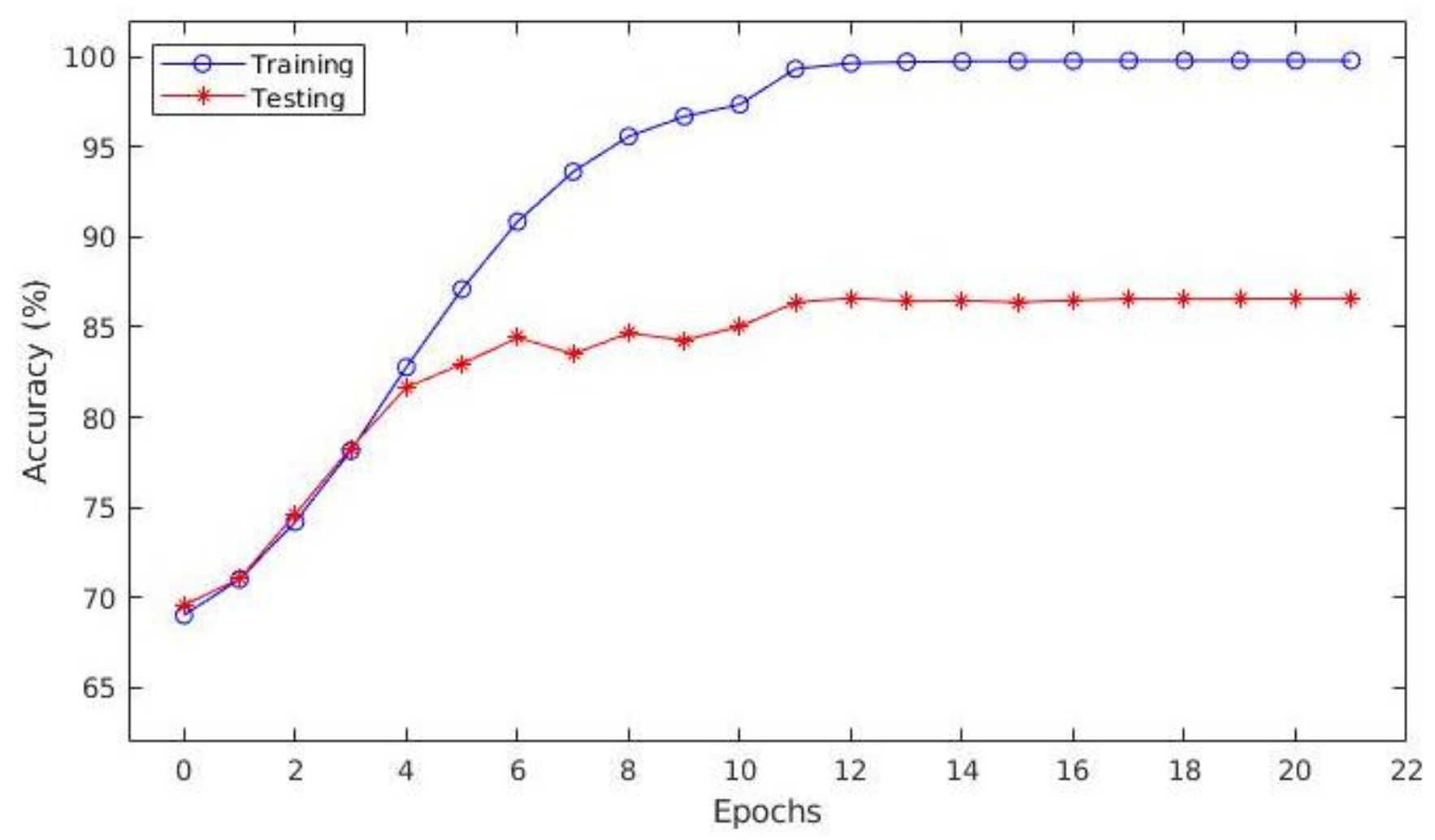}
        \caption{Training and validation accuracy} 
        \label{fig:training_epochs_accuracy}
    \end{subfigure}
  \begin{subfigure}[t]{0.5\textwidth}
        \centering
        \includegraphics[width=.75\textwidth]{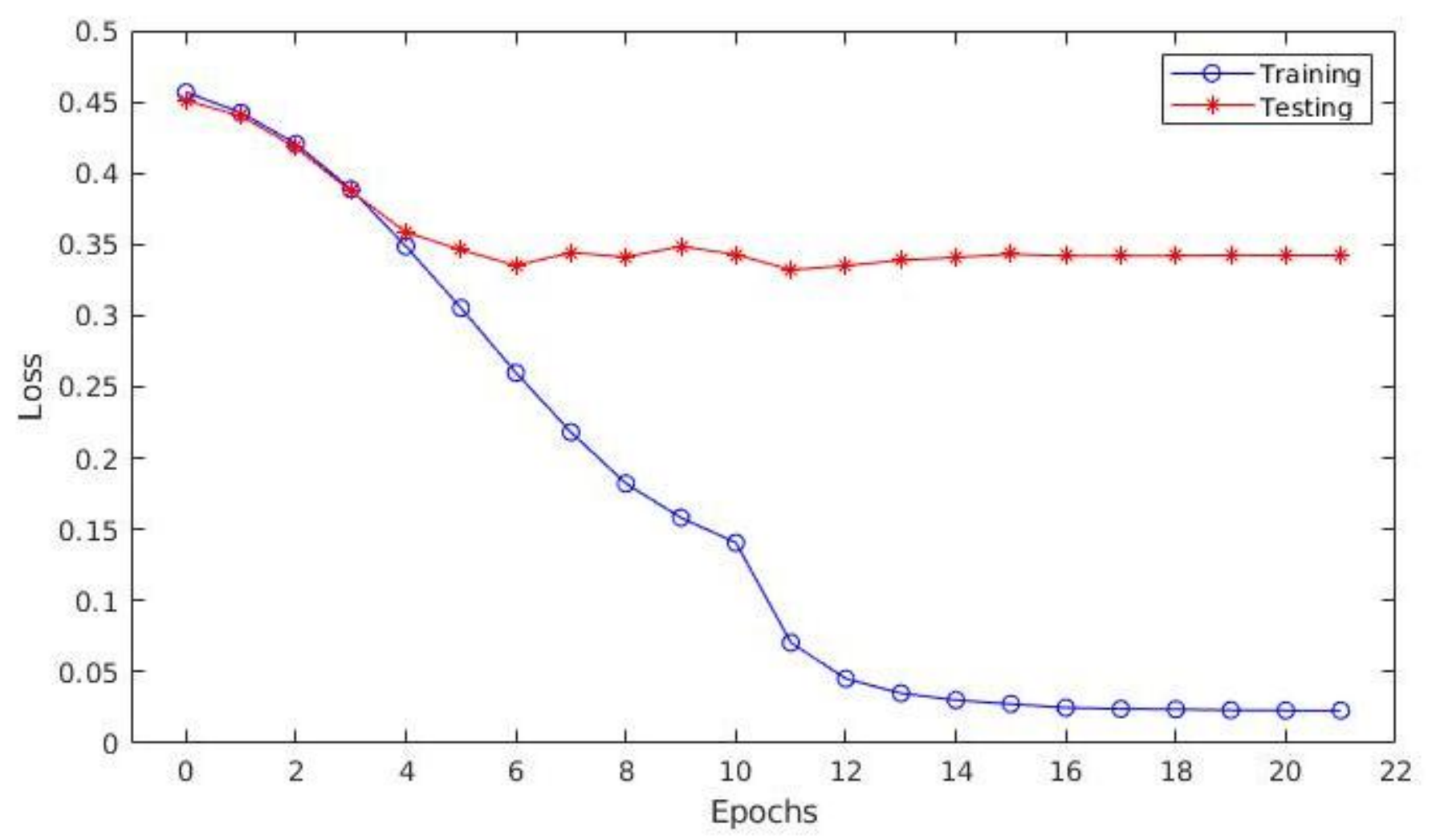}
        \caption{Training and validation loss} 
        \label{fig:training_epochs_loss}
    \end{subfigure}

  \caption{Training and validation loss and accuracy curves in terms of training epochs of AudVowelConsNet for PHQ-binary classes with an early stopping. %a) represent the training and validation accuracy of the network, while b) represents the training and validation loss of the network.
  } 
  \label{fig:perf_training_epochs-binary} 
\end{figure}

%\textcolor{red}{Figure 6 new place}
\begin{figure}[]
  \centering
  \begin{subfigure}[t]{0.5\textwidth}
        \centering
        \includegraphics[width=0.75\textwidth]{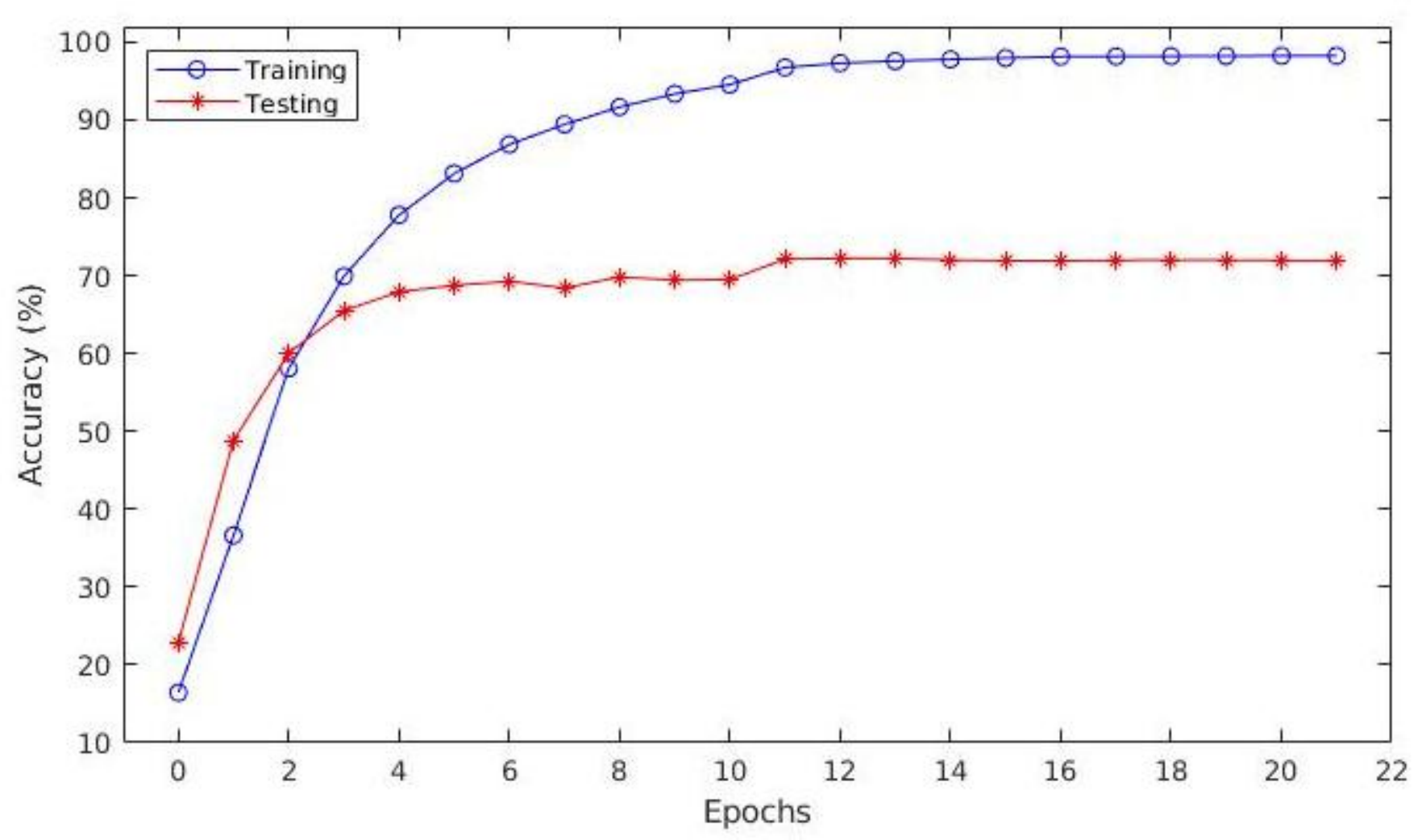}
        \caption{Training and validation accuracy. } 
        \label{fig:training_epochs_accuracy}
    \end{subfigure}
  \begin{subfigure}[t]{0.5\textwidth}
        \centering
        \includegraphics[width=.75\textwidth]{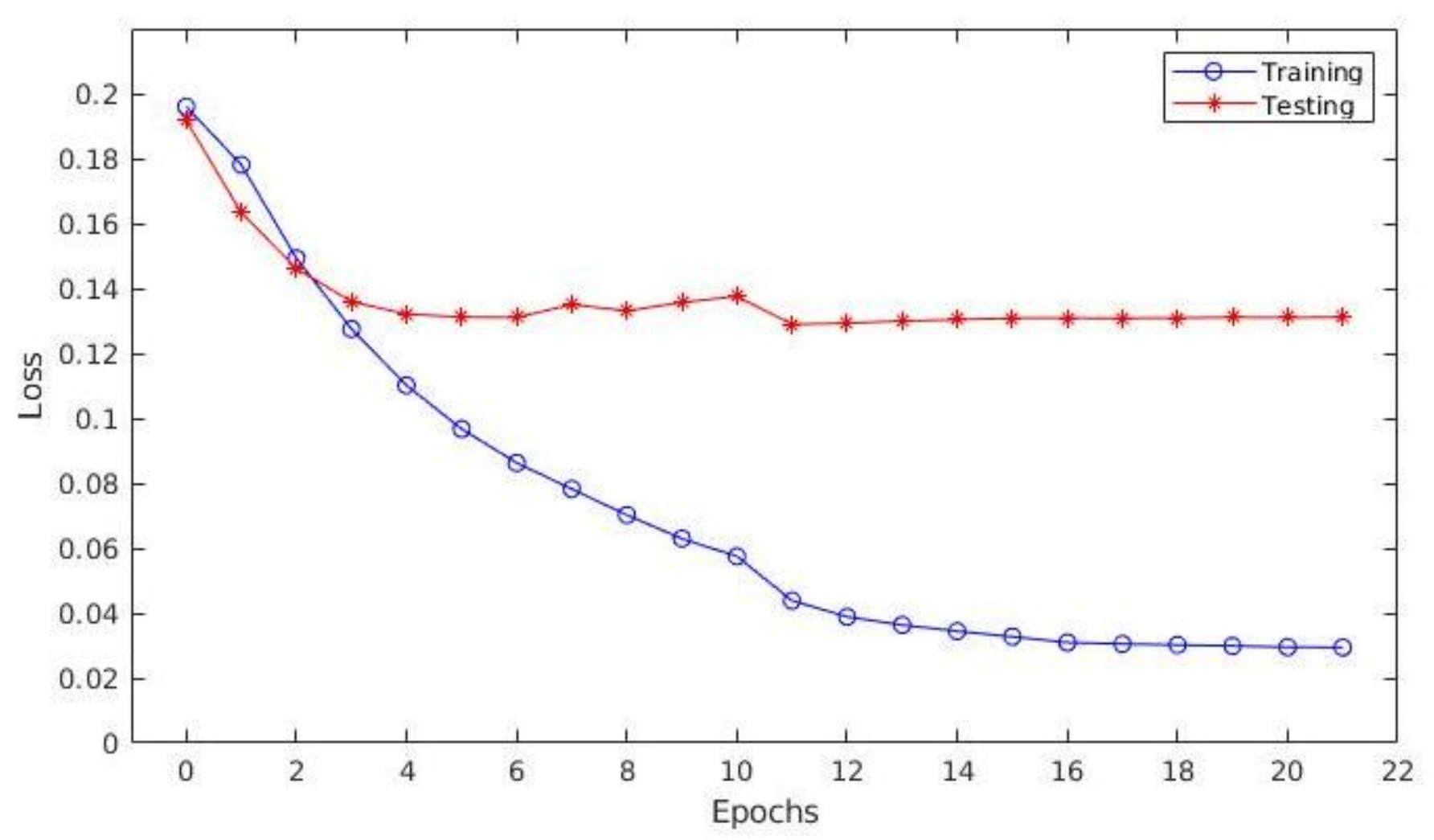}
        \caption{Training and validation loss.} 
        \label{fig:training_epochs_loss}
    \end{subfigure}

  \caption{Training and validation loss and accuracy curves in terms of training ephocs of AudVowelConsNet for PHQ-8 24 classes with an early stopping. %a) represent the training and validation accuracy of the network; While, b) represents the training and validation loss of the network.
  } 
  \label{fig:perf_training_epochs} 
\end{figure}

Furthermore, we present the AudVowelConsNet performances over the training epochs. Figure \ref{fig:perf_training_epochs-binary} represents the training and validation curves in terms of accuracy and loss for PHQ binary classes, while Figure \ref{fig:perf_training_epochs} shows the training and validation curves for PHQ-Score. 
A decrease in loss is observed for both experiments. After a certain number of epochs, the validation loss curves tend to flatten. Therefore, we perform early stopping in both experiments. Similarly, for both experiments, training and testing accuracy stopped increasing after a certain number of epochs. We notice few fluctuations in training and validation loss curves before the curves reach stability. This demonstrates that the AudVowelConsNet does not overfit at the end of training and validates its capacity to generalize to a new dataset.

%-----------------------------------------------------
\subsubsection{Comparison with state-of-the-art approaches}

%%===============Table 3=======================

% on depressed class its higher
\begin{table*}[pos=H,width=0.99\textwidth,align=\centering]
\caption{\label{tab_comparison_soa_binary} Comparison of the performances of the proposed deep neural networks with state of the art methods for predicting the PHQ-8 binary for depression recognition in terms of Precision, Recall and F-Score. (D) and (ND) are for Depression and Non-Depression classes, respectively. The table also summrizes the average F-score, Accuracy, RMSE and CC.}
\centering
%\begin{tabular}{|p{3.4cm}||p{1cm}|p{1cm}|p{1cm}|p{1cm}|p{1cm}|p{1cm}|p{1cm}|p{1.3cm}|p{1cm}|p{0.7cm}|}
\begin{tabular}{|l||c|c|c|c|c|c|c|c|c|c|}
\hline 
\textbf{Method} &\multicolumn{2}{c|}{ \textbf{Precision (\%)}} & \multicolumn{2}{c|}{\textbf{Recall (\%)}} & \multicolumn{3}{c|}{\textbf{F-Score (\%)}} & \textbf{Acc. (\%)} & \textbf{RMSE} & \textbf{CC} \\
\cline{2-8}
 & D &ND & D &ND & D &ND &Av.&&&\\
\hline \hline 
\cite{ma2016}  &  35 &100 & 100 &54 &--- & 70 & 52&---&---&--- \\
\cite{salekin2018} &--- &--- &--- & ---&---& ---&85.44 &\textbf{96.7}&--- &---\\
\cite{EmnaRejaibi2019}  &56.28 &79.48& 45.11 &85.85& 50 &83 & 75.39 &74.13 &0.47 &0.51 \\
\cite{rejaibi2019}  &  69 &78& 35 &94& 46 &85& 80.00 &76.27 &0.41&--- \\
\hline
%Audio Vowels Net &  67 &84& 66 &85&66 &85 & 78.99 &78.77& 0.46& 0.57 \\
%Audio Consonants Net &  73 &84& 64 &89&68 &86& 80.30 &80.98&0.44&0.62 \\
AudVowelConsNet &  \textbf{81} &\textbf{88} & \textbf{73} &\textbf{92}
&\textbf{77} & \textbf{90} & \textbf{85.85} &86.06&\textbf{0.37}&\textbf{0.72} \\
\hline 
\end{tabular}
\end{table*}

\textbf{PHQ-8 Binary -- }
Table \ref{tab_comparison_soa_binary} reports the Precision, Recall and F1-score of AudVowelConstNet  compared to deep learning-based state-of-the-art approaches of \cite{ma2016}, \cite{salekin2018}, \cite{EmnaRejaibi2019}, and \cite{rejaibi2019} for the Depression and Non-depression PHQ-8 binary classes. Additionally, the table summarizes the average F-score, Accuracy, RMSE and CC for this task.
We can see that AudioVowelConsNet obtained significantly higher values of Precision, Recall and F1-score for both classes (Depression and Non-Depression) as compared to existing state-of-the-art methods. 
For the Depression class, the model obtained  a precision of $81\%$, a recall of $73\%$ and an F1-score of $77\%$.
Compared to EmoAudioNet \citep{EmnaRejaibi2019} and MFCC-based RNN method of \citep{rejaibi2019}, AudioVowelConsNet predicts the Depression class more accurately increasing Precision by $43.92\%$ and $17.39\%$ respectively.  
Moreover, an increase of $10.72\%$ and $12.82\%$ in Non-Depression precision value is obtained with respect to \citep{EmnaRejaibi2019} and \citep{rejaibi2019}.
Compared to \cite{ma2016}, F1-score is increased by $65.1\%$ on average and by $28.57\%$ for Non-Depression class using the AudioVowelConsNet. 
Audio Vowels and Audio Consonants Nets achieved precision of $84\%$ for Non-Depression class and precision values of $67\%$ and $73\%$ respectively for Depression class outperforming the state of art methods (cf. Table \ref{tab_PHQ-8_binary_D_ND}).
\cite{salekin2018} reported an accuracy of $96.7\%$ and an F1-Score of $85.44\%$. AudVowelConsNet results in a lower accuracy ($86.06\%$) and a similar  F1-Score ($85.85\%$). This might be due to the leave-one-speaker out model evaluation strategy used by \cite{salekin2018}. Compared to the other proposed approaches where a simple train-test split strategy is adopted, training the  model on multiple train-test splits increases the model performance especially in small datasets. Similarly, for PHQ-binary an increment of $16.09\%$ and $12.84\%$ in accuracy is obtained as compared to EmoAudioNet \citep{EmnaRejaibi2019} and MFCC-based RNN method \citep{rejaibi2019}, respectively.

\textbf{PHQ-8 severity level -- }
Table \ref{tab_severity_level} summarizes state-of-the-art comparison results for predicting the depression severity levels in terms of RMSE. The proposed AudVowelConsNet performs better ($0.1429$) than the MFCC-based Recurrent Neural Network  architecture \citep{rejaibi2019} ($0.168$), EmoAudioNet (fusion of MFCC and Spectorgram based CNN Networks) \citep{EmnaRejaibi2019} ($0.18$). While the approach of \cite{yang2017b} has the best non-scaled RMSE of ($1.46$ for depressed male) for the PHQ-8 scores prediction. AudVowelConsNet comes in a close second with a non-normalized RMSE of $3.22$.

%\textcolor{red}{Table \ref{tab11} added by Muhammad Muzammel for missing values}
\begin{table}[]
\caption{\label{tab_severity_level} Severity level comparison of AudVowConsNet with existing state of the art methods for depression assessment. ($^{DM}$) : Depressed Male. ($^{Norm}$): Normalized RMSE } %\textcolor{cyan}{Muhamed, what is (Depressed Male). Depressed Male refers to the depression value for gender males only}}
\centering
\begin{tabular}{|l||l|}
\hline
\textbf{Method}  & \textbf{RMSE} \\
\hline \hline
\cite{valstar2016}  & 7.78 \\
\cite{yang2017}  & $5.59^{DM}$ \\
\cite{yang2017b} & $1.46^{DM}$  \\
\cite{EmnaRejaibi2019}   & $0.18^{Norm}$  \\
\cite{rejaibi2019}  & $0.17^{Norm}$  \\
\hline
AudVowelConsNet & \textbf{$0.14^{Norm}$} | \textbf{3.22} \\
%\cite{rejaibi2019}  & 0.1680  \\
%AudVowConsNet & \textbf{0.1429} \\
\hline
\end{tabular}
\end{table}

%\textcolor{red}{Figure 8 old place.}

\subsection{Discussion }

In this section we discuss the limitations and strengths of the proposed method with respect to the state of the art approaches. Moreover, we discuss the managerial implications of this approach for practice in psychiatry.
%Strength:

\subsubsection{Strengths and Limitations}
In this research, a new approach is presented to separate vowels and consonants from the speech and build a deep learning model combining high level features learned from such phoneme-level units for depression recognition. 
There are many existing models which detect clinical depression using speech. Yet, these models did not provide information about which portion of  speech  is mostly effected by the clinical depression. By analyzing consonants and vowels separately, the proposed approach  helped to understand the effect of depression on human pronunciation. Particularly, it allowed to understand which part of the speech (consonants or vowels) is more affected by depression. The results revealed that the speech consonants space is more affected in the depressed speech compared to the vowels space.    

Furthermore, most of the existing approaches extract features from the patient's full speech (i.e., including voiced and unvoiced segments). In this research, the removal of unwanted speechless segments reduces the noise in depression detection and leads to a better accuracy as compared to the existing state of the art approaches \citep{salekin2018,rejaibi2019,EmnaRejaibi2019}. 
%\textcolor{blue}{ please provide some references to back this statement up.}

Finally, the proposed AudVowConsNet can be applied for other speech-related applications. Extracting deep features from the vowels and from consonants separately could be a strong behavioural biomarker and a strong biometric trait for gender, ethnicity, dialect and emotion recognition. However, this statement merits further investigation. 

%Weaknesses:
%The separation/combination.

On the other hand, this research presents certain limitations. For instance, the proposed approach was limited to audio vowels and consonants based spectral features. While, many models extracted MFCC, spectral features and other features from the patient speech \citep{EmnaRejaibi2019,rejaibi2019}. Therefore, a further research can be performed to extract different acoustic features from acoustic vowels and consonants and compare the performance to the proposed model. %\textcolor{blue}{ This point is good and meaningful }  

%https://www.hindawi.com/journals/misy/2020/8454327/
Moreover, separating the speech signal to two separate phoneme units which are learned separately with two deep networks doubles the needed computational power. This makes the ubiquitous deployment of such model on resource limited platforms such as mobile or embedded platforms very challenging.

\subsubsection{Managerial Implications for Practice}
%\citep{su2020deep}
The proposed approach meets the growing demand for the development of computer-aided diagnosis systems and artificial medical assistants using AI techniques in psychiatry.
 
Such AI-based systems can assist clinicians in making more accurate predictions of mental illness diagnosis and complement the potential misdiagnosis of interview-based diagnosis practices conventionally used by such health practitioners. An essential requirement for the  development of reliable AI-based systems is to detect propitious behavioural biomarkers capable of reflecting pathological attributes of psychiatric disorders.  

The proposed approach have the potential to be  applied for real-world applications in several forms including artificial medical assistants, web applications, computer-aided diagnosis systems, etc. Such applications, can be integrated in the treatment plan allowing to complement conventional  therapies by facilitating:

\begin{itemize}
\item Patient monitoring: monitor patient's vocal activity and prevent the onset of a mental health crisis by seeking the health practitioner's aid at the right time or by using support groups.
\item Interventions recommendations: detect depressive episodes and recommend personalized interventions to assist users in maintaining an emotional balance.
\end{itemize}

\section{Conclusion and Future Work}
\label{sec5}

In this paper, we propose an Artificial Intelligence (AI) based application for clinical depression recognition and assessment from speech. A deep phoneme-level analysis and learning is performed and three spectrogram-based CNNs networks that rely on  vowel, consonant spaces and fusion of both are studied. Consonant based CNN architecture outperforms the vowel CNN network. The fusion of both networks improves significantly the results and a notable increment in performance is observed. Further, we compare the fusion network (named AudVowelConsNet) with state of the art benchmark approaches on the DAIC-WOZ dataset. Our model outperforms all existing state-of-the-art approaches in depression recognition. These results show that the proposed deep learning based analysis of vowels and consonants is highly reliable in automatic diagnosis of clinical depression. 
In future work, we aim to verify these findings on other depression-speech databases. 
Moreover, we are interested in studying the fusion of different modalities such as  audio, textual, and  visual modalities for more accurate AI-based application for clinical depression recognition.
As a matter of fact, depressed individuals tend to speak slowly and use simple phrases in their speech, compared to non-depressed individuals which express their positive feelings in speech. Therefore, exploring techniques for automatic speech transcription and the extraction of discriminative linguistic features to complement acoustic based features present a future direction for this work.

Furthermore, as the aim of this research was to investigate the contributions of vowels and consonants in depression recognition, unvoiced segments were removed from the patient speech. The removal of these unvoiced segments improved the performance of proposed model. Yet, depressed individuals tend to have more unvoiced segments as compared to voiced segments in speech. In the future, we plan to investigate the integration of the silence duration as a feature within the proposed model.

\bibliographystyle{apa}
\bibliography{refs}

%%%%%%%%%%%%%%%%%%%%%%%%%%%%%%%%%%%%%%%%%%%%%%

%\bibliography{mybibfile}

\end{document}

%% file: Graphics/PHQ_binary_dispersion.tex
\def\angle{0}
\def\radius{3}
\def\cyclelist{{"orange","blue","red","green"}}
\newcount\cyclecount \cyclecount=-1
\newcount\ind \ind=-1
\begin{tikzpicture}[nodes = {font=\sffamily}]
  \foreach \percent/\name in {
      29.83/Depressed,
      70.17/Non-Depressed 
    } {
      \ifx\percent\empty\else               % If \percent is empty, do nothing
        \global\advance\cyclecount by 1     % Advance cyclecount
        \global\advance\ind by 1            % Advance list index
        \ifnum3<\cyclecount                 % If cyclecount is larger than list
          \global\cyclecount=0              %   reset cyclecount and
          \global\ind=0                     %   reset list index
        \fi
        \pgfmathparse{\cyclelist[\the\ind]} % Get color from cycle list
        \edef\color{\pgfmathresult}         %   and store as \color
        % Draw angle and set labels
        \draw[fill={\color!50},draw={\color}] (0,0) -- (\angle:\radius)
          arc (\angle:\angle+\percent*3.6:\radius) -- cycle;
        \node at (\angle+0.5*\percent*3.6:0.7*\radius) {\percent\,\%};
        \node[pin=\angle+0.5*\percent*3.6:\name]
          at (\angle+0.5*\percent*3.6:\radius) {};
        \pgfmathparse{\angle+\percent*3.6}  % Advance angle
        \xdef\angle{\pgfmathresult}         %   and store in \angle
      \fi
    };
\end{tikzpicture}

%% file: Graphics/PHQ_scores_dispersion_gender.tex
\begin{tikzpicture}
  \centering
  \begin{axis}[
        ybar, axis on top,
        height=4cm, width=20cm,
        bar width=0.25cm,
        ymajorgrids, tick align=inside,
        %major grid style={draw=white},
        enlarge y limits={value=.1,upper},
        enlarge x limits=0.02,
        ymin=0, ymax=20,
        axis x line*=bottom,
        axis y line*=left,
        y axis line style={opacity=0},
        tickwidth=0pt,
        legend style={
            at={(0.5,-0.2)},
            anchor=north,
            legend columns=-1,
            /tikz/every even column/.append style={column sep=0.5cm}
        },
        symbolic x coords={0,1,2,3,4,5,6,7,8,9,10,11,12,13,14,15,16,17,18,19,20,21,22,23},
       xtick=data,
       nodes near coords={
        \pgfmathprintnumber[precision=0]{\pgfplotspointmeta}
       }
    ]
    \addplot coordinates {(0,12) (1,8) (2,6) (3,6) (4,5) (5,2) (6,3) (7,4) (8,2) (9,6) (10,5) (11,5) (12,4) (13,1) (14,0) (15,2) (16,2) (17,1) (18,2) (19,1) (20,3) (21,1) (22,1) (23,1)};
\addplot coordinates {(0,13) (1,9) (2,11) (3,9) (4,5) (5,8) (6,3) (7,9) (8,3) (9,2) (10,5) (11,2) (12,3) (13,2) (14,1) (15,3) (16,3) (17,3) (18,2) (19,1) (20,1) (21,0) (22,0) (23,0)};
\legend{Female,Male}
  \end{axis}
  \end{tikzpicture}